# Hydrodynamics of chiral nematics in a channel and sudden contraction geometry FREE

Isreal Morawo ; Dana Grecov

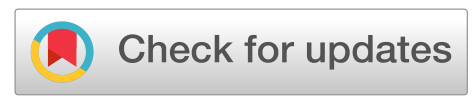

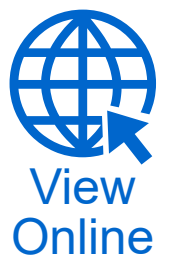
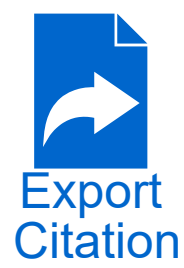

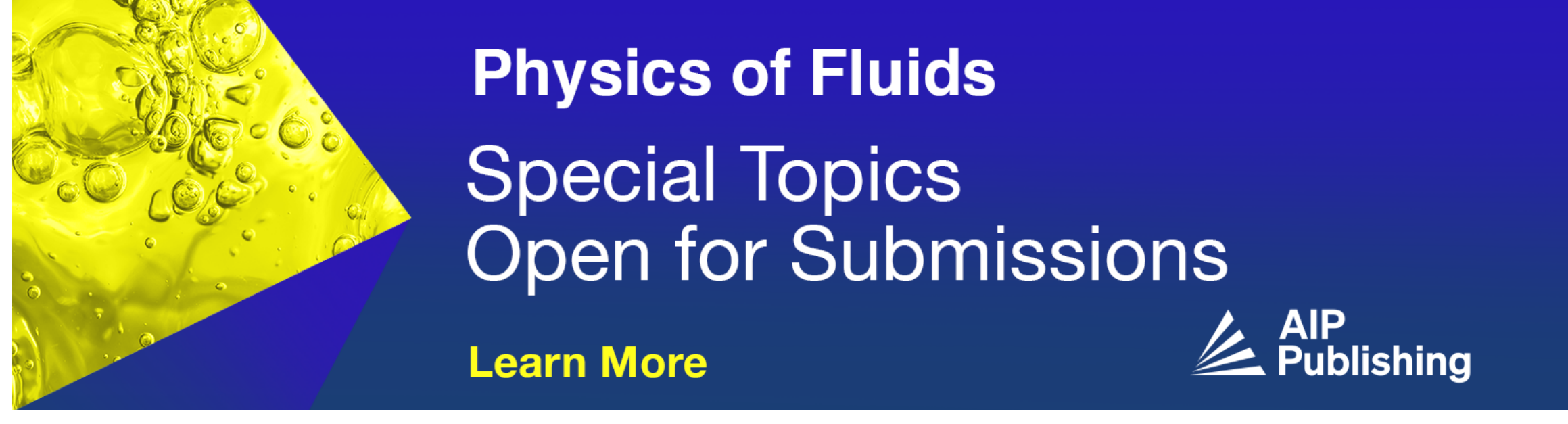

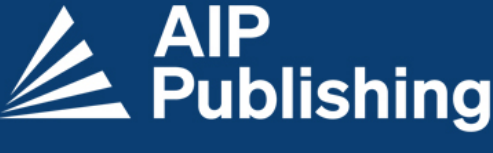

# Hydrodynamics of chiral nematics in a channel and sudden contraction geometry

View Online

Export Citation

CrossMark

Isreal Morawo 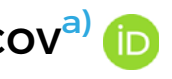 and Dana Grecov[a)]

## AFFILIATIONS

Department of Mechanical Engineering, University of British Columbia, Vancouver, British Columbia V6T 1Z4, Canada

[a)]Author to whom correspondence should be addressed: dgrecov@mech.ubc.ca

## ABSTRACT

This study investigates the influence of chirality, viscous effects, and confinement geometry on the flow dynamics and defect structures of cholesteric liquid crystals (CLCs) using numerical simulations. As chiral strength increases, $\pi$-twist defects form at low chirality and progressively organize into hexagonal domains resembling blue phase-like structures at higher chirality. At sufficiently high chirality and aspect ratios, localized defect arrangements emerge that are structurally analogous to half-skyrmion configurations observed in confined cholesteric systems. Although the skyrmion number is not explicitly calculated, the defect arrangement mirrors known equilibrium patterns. These results indicate dynamic analogs of equilibrium phases formed and stabilized by flow–structure interactions. Sudden contraction geometries reveal how aspect ratio and elastic interactions promote flow disturbances and defect development. Analyses of velocity and scalar order parameter ($S$) distributions demonstrate strong coupling between molecular alignment and hydrodynamic fields, with localized vortices forming around $\tau^-$ defects in regions of low $S$ and velocity. Temporal evaluations of velocity fluctuations uncover irregular or quasi-chaotic flow regimes at low Ericksen numbers (Er), characterized by irregular defect motion and skewed probability density functions. These findings offer new insight into CLC structure formation and flow behavior, with potential applications in defect engineering, microfluidics, soft robotics, and photonic materials.

## I. INTRODUCTION

Liquid crystals (LCs) represent a unique state of matter that displays properties different from conventional liquids and solids. Their mechanical properties, such as the ability to flow like a liquid while having elasticity like solids (viscoelasticity) and varying properties with direction (anisotropy), along with symmetry characteristics like orientational order of molecules and sometimes partial positional order, classify them as mesophase, an intermediate form.[1] LCs have a range of unique structures depending on their molecular arrangement and environmental conditions, with three main types being nematic, smectic, and cholesteric phases.

Cholesteric liquid crystals (CLCs) exhibit a helical twist, defined by the helical pitch ($P_0$) and a twist sense.[2,3] The twist sense describes the direction of their helical arrangement, determined by molecular chirality. A left-handed helix twists counterclockwise along the axis, while a right-handed helix twists clockwise. This chirality-induced helical structure defines key properties of CLCs, including selective light reflection. Under high chirality, these LCs can give rise to blue phases (BPs), characterized by three-dimensional cubic structures with regions of double-twist.[4–6] In two dimensions, these phases form hexagonal lattices, often referred to as merons or half-skyrmions.[7] Their unique and diverse structures enable a wide range of applications, from engineered systems to biological materials, such as DNA, silk, and chitin.[8] Notably, these materials have become a preferred choice for drug delivery, as they improve drug solubility while allowing precise control over release rates.[9,10]

CLCs can be formed from various materials, including small organic molecules and chiral polymers that self-organize into helical structures, resulting in their characteristic optical properties. Common examples include LC monomers and oligomers, which can be used for specific thermal and optical behaviors.[11,12] In addition to these traditional materials, nanomaterials, like cellulose nanocrystals (CNCs) can also form CLC phases under specific conditions.[13–17] CNCs self-assemble in suspension due to their rod-like structure and, in a certain range of concentrations, form CLCs. Esmaeili *et al.*[18] explored how the initial structure of CLC suspensions, which form CNCs, impacts the alignment during the 3D printing under shear flow conditions. Long and Selinger[19] examined CLCs confined between two infinite parallel plates with free boundary conditions, observing that double twists arranged in a row, with semicircular domains at the surface increasing in size as the natural pitch grows.

Despite their well-defined helical structure, CLCs can exhibit distortion leading to defect formation, which influences material properties and optical behavior.[1] Dislocations represent line defects in the helical structure of CLCs, where the regular helical arrangement is interrupted, leading to structural irregularities.[2] These defects are visible under polarizing microscopy and have distinct optical properties, making them essential for understanding the optical behavior of CLCs.[20] Light-responsive control of cholesteric structures has also been demonstrated in experiments,[21] highlighting the broad tunability of these systems for photonic applications. According to Araki *et al.*,[22] strong geometrical frustrations between nematic ordering and anchoring in LCs can result in the formation of stable topological defects. These defects, stabilized by topological constraints from solid surfaces, offer new possibilities for defect engineering in LCs. There are three types of defects common to CLCs, namely, $\tau^{(\pm s)}$, $\chi^{(\pm s)}$, and $\lambda^{(\pm s)}$.[23] In $\lambda$ defects, the director field is continuous, but the pitch axis twists and flips to the third dimension at the defect line. $\tau$ defects are true disclinations, where the director drops at the core, leading to lower-$S$. $\chi$ defects typically occur in 3D geometry, acting like dislocations, and are, thus, not discussed in this article.

Theoretical modeling plays a crucial role in understanding the behavior of LCs. Several frameworks, including the Landau–de Gennes (LdG) theory,[5,15,24–29] the Leslie–Ericksen theory,[30–32] the Beris–Edwards model,[33] mean-field theory,[34] elastic continuum theory,[1] nonlinear dynamics approaches,[2,35] as well as molecular dynamics and Monte Carlo simulations.[36] Among these models, the tensorial LdG framework has proven particularly effective in describing microstructural behavior and defect dynamics.[1] This approach captures both short-range and long-range elasticities through the second-order tensor order parameter **Q**. LdG-based models have been widely applied in simulating defect nucleation[37] and LC flow when coupled with the Navier–Stokes (NS) equations, incorporating anisotropic and elastic stress contributions.

Numerical studies have advanced our understanding of CLC behavior under flow. Noroozi *et al.*[16] used numerical simulations of the LdG equations to study transient simple shear flow between parallel plates of chiral LCs, comparing their findings with experimental rheological data and observing cholesteric patterns at low shear rates and a flow-aligning regime at higher shear rates. Esmaeili *et al.*[18] employed hydrodynamic simulations based on LdG to investigate chiral structures, and their study revealed that these structures transform into complex twisted configurations at low flow velocities corresponding to low Ericksen Number (Er). These simulation results are consistent with observations of flow-induced birefringence patterns in CNC suspensions, where increasing the flow rate leads to distinct alterations in iridescent patterns. This *in situ* characterization offered valuable insights into the structural evolution of anisotropic and chiral CNC alignment during the direct ink writing process of CNC-based responsive materials. Venhaus *et al.*[38] used LdG theory to simulate the alignment dynamics of CLCs under flow, overcoming previous limitations related to helicity direction and pre-imposed pitch, and developed phase diagrams to analyze the system's behavior under varying chiral and flow forces. Hicks and Walker[24] propose a fully implicit $L^2$ gradient flow for computing energy minimizers of the LdG model, identifying a time step and mesh size restriction to avoid numerical artifacts, particularly when simulating CLCs.

Research on CLC flow in confined geometries has highlighted the effects of viscous and chiral interactions. Li and Grecov[25] found that in a lid-driven cavity, increasing viscous effects disrupted the chiral structure and reduced the number of defects, while Nikzad and Grecov[15] showed that at lower De, a hexagonal structure formed due to the dominance of chirality, and higher De resulted in the disappearance of this structure. In the context of sudden contractions, Fedorowicz and Prosser[33] revealed that geometry significantly affects the flow behavior and microstructure of LCs, impacting material performance and pressure loss predictions. Cruz *et al.*[30] observed that increasing viscosity reduces corner vortex size in viscoelastic LC flows through contractions, while director alignment remained consistent across different Er.

Sudden contractions are crucial in various engineering and biological systems, influencing fluid transport, mixing, and stability. In the context of viscoelastic fluids, these geometric features can give rise to complex flow patterns, making their study essential for both fundamental understanding and practical applications. Galeel *et al.*[39] observed that increasing the Deborah number (De) raises stresses and reduces recirculation zones in sudden contractions of a viscoelastic fluid. Similarly, Fu *et al.*[40] noted that viscoelasticity minimizes the extent and intensity of recirculation zones in expansion channels, while Xue *et al.*[41] and Luo[42] observed the expansion of recirculation zones in planar contractions. Raihan *et al.*[43] emphasized the significant role of confinement in determining extensional flow instabilities within single cavity microchannels. Additionally, Wu *et al.*[44] highlighted how polymer elasticity influences flow regimes and vortex development in sudden contractions and expansions. These findings indicate that viscoelastic effects can dampen vortex strength and size in confined geometries.[45–48] In viscoelastic fluids, elasticity can induce irregular flow patterns, a phenomenon known as elastic turbulence, characterized by velocity fluctuations driven by elastic stresses rather than inertia. This has been observed in both polymeric materials and active LCs.[49–52]

This paper investigates the behavior of CLCs within channels and sudden contraction geometries. By examining their response to flow dynamics within these geometries, this research provides insights into their behavior under different flow conditions. This study utilizes the LdG theory and the NS with a modified stress tensor. These findings deepen our understanding of CLC flow behavior, with practical implications for advanced manufacturing applications, such as 3D printing and additive manufacturing, where controlling the pattern of the microstructure and reducing the number of defects is important. Additionally, the study highlights the potential of CLCs in drug delivery systems, where their orientation and microstructural changes can be leveraged to control the release and distribution of therapeutic agents within. Understanding the effect of chirality, viscous effects, on the coupling between flow and structure is essential for developing predictive models that optimize structural organization and minimize control defect numbers formation, which is important for optical applications where the pitch influences selective reflection, and defects impact light scattering.

The paper is structured as follows: The introduction provides an overview of LCs and their unique properties. The methodology outlines the theoretical framework, focusing on the LdG equations and their coupling with the NS equations to simulate CLC flow. Results and discussions present the simulation outcomes, analyzing molecular alignment, defect formation, and flow for both channels and sudden contractions. Finally, the conclusions summarize the study's contributions, implications for practical applications, and directions for future research.

## II. THEORY AND GOVERNING EQUATIONS

In the LdG theory, the microstructure of LCs is effectively characterized by a second-order, symmetric, and traceless tensor order parameter, $\mathbf{Q}$[2]:

$$\mathbf{Q} = \int \left( \boldsymbol{\vartheta}\boldsymbol{\vartheta} - \frac{1}{3}\mathbf{I} \right) \omega(\boldsymbol{\vartheta})\, d^2\boldsymbol{\vartheta}, \tag{1}$$

$$\mathbf{Q} = S\left( \mathbf{nn} - \frac{1}{3}\mathbf{I} \right) + \frac{1}{3}P(\mathbf{mm} - \mathbf{ll}), \tag{2}$$

where

$$S = \frac{3}{2}\mu_n \quad \text{and} \quad P = 3\left( \mu_n + \frac{3}{2}\mu_m \right).$$

Here, $\boldsymbol{\vartheta}$ denotes a unit vector representing the LC molecule's orientation while $\omega$ is the orientation distribution function (ODF). The ODF quantifies the statistical distribution of molecular orientations within the LC system, influencing the material's properties. $S$ is the scalar order parameter, which illustrates the degree of alignment, and it is defined as $\langle P_2 \cos\theta \rangle$ where $P_2$ is the second Legendre polynomial. The values of $S$ range from isotropy (0) to perfect alignment (1).[25,53,54]

Finally, the equilibrium scalar order parameter $S_{eq}$ is expressed as

$$S_{eq} = 0.25 + 0.75\sqrt{1 - \frac{8}{3\varkappa}}, \tag{3}$$

where $\varkappa$ is the nematic potential.[16,53]

Additionally, $P$ represents the biaxial order parameter, although this paper focuses on uniaxial cholesteric and nematic liquid crystals (CLCs and LCs), so $P$ is close to zero.

Typically, the complete three-dimensional $\mathbf{Q}$ is represented by a $3 \times 3$ matrix with components[2]

$$\mathbf{Q} = \begin{pmatrix} Q_{11} & Q_{12} & Q_{13} \\ Q_{21} & Q_{22} & Q_{23} \\ Q_{31} & Q_{32} & -(Q_{11} + Q_{22}) \end{pmatrix}. \tag{4}$$

The traceless and symmetric nature of $\mathbf{Q}$ limits its independent components to five.

### A. Mathematical modeling of CLCs

The evolution equation describing the flow of CLCs is given by[38,53]

$$\overline{\mathbf{Q}} = F(\nabla u, \mathbf{Q}) + H^{sr}((De)^{-}, \mathbf{Q}) + H^{lr}((De), \nabla\mathbf{Q}) + \mathbf{P}(\mathbf{Q}), \tag{5}$$

where $\overline{\mathbf{Q}}$ is the Jaumann derivative of $\mathbf{Q}$, $\mathbf{F}$ is the flow contribution, $\mathbf{H}^{sr}$ represents the short-range elastic contribution, $\mathbf{H}^{lr}$ represents the long-range elastic contribution, and $\mathbf{P}$ represents the chiral contribution.[53]

The flow contribution $\mathbf{F}$ is given by

$$F = \frac{2}{3}\beta\mathbf{A} + \beta\left( \mathbf{A}\cdot\mathbf{Q} + \mathbf{Q}\cdot\mathbf{A} - \frac{2}{3}(\mathbf{A}:\mathbf{Q})\,\mathbf{I} \right) - \frac{1}{2}\beta\,(\mathbf{A}\cdot\mathbf{Q})\cdot\mathbf{Q} + (\mathbf{A}\cdot\mathbf{Q}\cdot\mathbf{Q}) + (\mathbf{Q}\cdot\mathbf{A}\cdot\mathbf{Q}) + (\mathbf{Q}\cdot\mathbf{Q}\cdot\mathbf{A}) - ((\mathbf{Q}\cdot\mathbf{Q}):\mathbf{A})\mathbf{I}, \tag{6}$$

where $\beta$ is the shape parameter, which characterizes the shape of the CLC system under consideration. The term $\mathbf{A}$ denotes the rate of deformation.

The chiral term is defined by

$$P_{ij} = \sum_{i=1}^{3}\sum_{j=1}^{3}\sum_{k=1}^{3}\left( \varepsilon_{mik}\frac{\partial Q_{mj}}{\partial x_k} + \varepsilon_{mjk}\frac{\partial Q_{mi}}{\partial x_k} \right), \tag{7}$$

where $\varepsilon_{ijk}$ denotes the Levi-Civita symbol, and the chiral term $\mathbf{P}$ contributes to the formation of director twisting patterns, which compete with the elasticity terms. $x_k$ denotes the spatial coordinates $(x, y, z)$.

$$\mathbf{H}^{sr} = 6\,\overline{D_r}\left( \frac{\varkappa}{3} - 1 \right)\mathbf{Q} + \varkappa\,\mathbf{Q}\cdot\mathbf{Q} - \varkappa\,(\mathbf{Q}:\mathbf{Q})\left( \mathbf{Q} + \frac{1}{3}\mathbf{I} \right). \tag{8}$$

$\mathbf{H}^{sr}$ corresponds to the short-range elasticity, it arises directly from the intermolecular attractive and repulsive forces, such as van der Waals forces, excluding the volume effect. This, in turn, controls the isotropic-nematic transition.[53]

$$\overline{D_r}(\mathbf{Q}) = \frac{D_r}{\left( 1 - \frac{3}{2}(\mathbf{Q}:\mathbf{Q}) \right)^2}, \tag{9}$$

where $\overline{D_r}$ is the microstructural rotational diffusivity, which quantifies the rate of rotational motion of molecules within the CLCs.[54]

$\mathbf{H}^{lr}$ stands for the long-range elasticity, also known as the *Frank elasticity*. It characterizes the secondary effects of nematic intermolecular forces, transmitting surface anchoring effects from the boundaries into the material.[53,55]

$$\mathbf{H}^{lr} = \frac{6\,\overline{D_r}\,L_1}{ck_BT}\left( \nabla^2\mathbf{Q} + \frac{L^*}{2}\left(\nabla(\nabla\cdot\mathbf{Q}) + (\nabla(\nabla\cdot\mathbf{Q}))^T\right) - \frac{2}{3}\,\mathrm{tr}(\nabla(\nabla\cdot\mathbf{Q})) \right) \tag{10}$$

where $L^* = \frac{L_2}{L_1}$ and $L_1$ and $L_2$ are the Landau elastic coefficients, defined as $L_1 = \frac{K_{22}}{2S^2}$ and $L_2 = \frac{K_{11} - K_{22}}{S^2}$; $K_{ii}$ are the Frank elasticity constants, which depend on the principal deformation of the molecules, $c$ is the molecular concentration, $k_B$ is the Boltzmann constant, and $T$ is the absolute temperature.

$\overline{\mathbf{Q}}$ is the Jaumann derivative of $\mathbf{Q}$ and $\mathbf{W}$ is the vorticity tensor[16]

$$\overline{\mathbf{Q}} = \frac{\partial\mathbf{Q}}{\partial t} + (\mathbf{u}\cdot\nabla)\mathbf{Q} - \mathbf{W}\cdot\mathbf{Q} + \mathbf{Q}\cdot\mathbf{W}. \tag{11}$$

The Ericksen number ($Er$), Deborah number ($De$), Chiral strength ($\Theta$), Energy Ratio ($R$), and Reynolds number ($Re_n$) are the dimensionless parameters used in this paper. $Er$ quantifies the relative impact of viscous flow compared to long-range elasticity effects. $De$ is the ratio of the internal timescale to the flow timescale. $R$ provides insight into the CLC system balance between short-range and long-range elasticities. $Re_n$ is a new Re that characterizes the balance between the inertial forces and the viscous forces for the CLC system, with $\rho$ representing the characteristic density.[15,25,56]

$$Er = \frac{ck_BT\,\dot{\gamma}\,L^2}{6L_1D_r}, \quad De = \frac{\dot{\gamma}}{6D_r}, \quad R = \frac{ck_BTL^2}{L_1}, \quad Re_n = \frac{\rho U^2}{ck_BT}. \tag{12}$$

Here, $\dot{\gamma}$ represents the shear rate, $U$ is the velocity, and $L$ represents the characteristic length scale of the flow.

The incompressible continuity equation and the momentum equations are presented as follows:

$$\nabla \cdot \mathbf{u} = 0, \tag{13}$$

$$\frac{D\mathbf{u}}{Dt} = \frac{1}{\rho}(-\nabla p + \nabla \cdot \boldsymbol{\tau}_t). \tag{14}$$

Here, $\mathbf{u}$ represents the velocity vector and $p$ denotes the pressure.

The total stress tensor $\boldsymbol{\tau}_t$ comprises several components, each essential for describing the system's behavior. $\boldsymbol{\tau}_v$ represents the viscous contribution, and $\boldsymbol{\tau}_e$ denotes the elastic contribution. Their combination yields the symmetric part of the total stress tensor, denoted as $\boldsymbol{\tau}_s = \boldsymbol{\tau}_v + \boldsymbol{\tau}_e$. Apart from the symmetric contributions, the total stress tensor encompasses an asymmetric component, $\boldsymbol{\tau}_a$, which accounts for deviations from ideal symmetric behavior within the system. Finally, the total stress tensor also includes the Ericksen stress tensor, $\boldsymbol{\tau}_{\mathrm{Er}}$, which describes the response of the CLC system to external shear forces. It represents the degree of orientation of the CLC molecules in the flow.[56]

$$\boldsymbol{\tau}_t = \boldsymbol{\tau}_a + \boldsymbol{\tau}_s + \boldsymbol{\tau}_{Er}, \tag{15}$$

$$\begin{aligned}\boldsymbol{\tau}_v = {} & \nu_1 \mathbf{A} + \nu_2\left(\mathbf{Q}\cdot\mathbf{A} + \mathbf{A}\cdot\mathbf{Q} - \frac{2}{3}(\mathbf{Q}:\mathbf{A})\mathbf{I}\right) \\ & + \nu_4\big((\mathbf{A}:\mathbf{Q})\mathbf{Q} + \mathbf{A}\cdot\mathbf{Q}\cdot\mathbf{Q} + \mathbf{Q}\cdot\mathbf{A}\cdot\mathbf{Q} \\ & + \mathbf{Q}\cdot\mathbf{Q}\cdot\mathbf{A} + (\mathbf{Q}:\mathbf{Q})\mathbf{I}\big)\end{aligned} \tag{16}$$

$$\begin{aligned}\boldsymbol{\tau}_e = ck_B T\Bigg[ & \frac{2\beta}{3}\mathbf{H} - \beta\left(\mathbf{H}\cdot\mathbf{Q} + \mathbf{Q}\cdot\mathbf{H} - \frac{2}{3}(\mathbf{Q}:\mathbf{H})\mathbf{I}\right) \\ & + \frac{\beta}{2}\big((\mathbf{H}:\mathbf{Q})\mathbf{Q} + \mathbf{H}\cdot\mathbf{Q}\cdot\mathbf{Q} + \mathbf{Q}\cdot\mathbf{H}\cdot\mathbf{Q} \\ & + \mathbf{Q}\cdot\mathbf{Q}\cdot\mathbf{H} + (\mathbf{Q}:\mathbf{Q})\mathbf{I}\big)\Bigg]\end{aligned} \tag{17}$$

$$\boldsymbol{\tau}_a = ck_B T(\mathbf{H}\cdot\mathbf{Q} - \mathbf{Q}\cdot\mathbf{H}), \tag{18}$$

$$\boldsymbol{\tau}_{Er} = ck_B T L_1\left(L^*(\nabla\cdot\mathbf{Q})\cdot(\nabla\mathbf{Q})^T - \nabla\mathbf{Q}:(\nabla\mathbf{Q})^T\right), \tag{19}$$

where $\mathbf{H} = \mathbf{H}^{\mathrm{sr}} + \mathbf{H}^{\mathrm{lr}}$, and $\nu$'s are the Landau viscosity coefficients.

Due to their complexities, CLCs exhibit multiple scales of time and length. The corresponding symbols for these time scales and length scales are provided in Refs. 25 and 53.

The dimensionless forms of the equations of conservation of mass and momentum are expressed as

$$\nabla\cdot\mathbf{u}^* = 0, \tag{20}$$

$$\frac{D\mathbf{u}^*}{Dt^*} = \frac{1}{Re_n}\left(-\nabla p^* + \nabla\cdot\boldsymbol{\tau}_t^*\right), \tag{21}$$

$$\begin{aligned}\frac{\partial\mathbf{u}^*}{\partial t^*} + (\mathbf{u}^*\cdot\nabla^*)\mathbf{u}^* = {} & \frac{1}{Re_n}\Big(-\nabla^* p^* + De(\nabla^*\cdot\boldsymbol{\tau}_v^*) \\ & + \nabla^*\cdot\boldsymbol{\tau}_e^* + \nabla^*\cdot\boldsymbol{\tau}_a^*\Big) + \frac{1}{R}\nabla^*\cdot\boldsymbol{\tau}_{Er}^*\end{aligned} \tag{22}$$

The dimensionless evolution equation is given by

$$\overline{\mathbf{Q}} = F^* + \frac{1}{De}\mathbf{H}^{\mathrm{sr}*} + \frac{1}{Er}\mathbf{H}^{\mathrm{lr}*} + \frac{\Theta}{Er}\mathbf{P}^*. \tag{23}$$

$\Theta$ describes the chiral strength, which is a dimensionless parameter for CLC, defined as a dimensionless quantity that quantifies the ratio of anisotropic distortional elasticity to isotropic elasticity. It is expressed in terms of the isotropic elastic length scale $L$ and the anisotropic chiral length scale $\mathscr{L}$,[15,16,25,38,57,58]

$$\Theta = \frac{L^2}{\mathscr{L}^2}. \tag{24}$$

The chiral term is given by

$$P_{ij}^* = \sum_{i=1}^{3}\sum_{j=1}^{3}\sum_{k=1}^{3}\left(\varepsilon_{mik}\frac{\partial\mathbf{Q}_{mj}}{\partial x_k^*} + \varepsilon_{mjk}\frac{\partial\mathbf{Q}_{mi}}{\partial x_k^*}\right). \tag{25}$$

## III. NUMERICAL SETUP

COMSOL Multiphysics, a finite element solver, was used to solve a system of five nonlinear equations representing the microstructure evolution process. The total stress derived from the evolution equations was incorporated into the Navier–Stokes (NS) equations. These equations were implemented using both the *Laminar Flow* module and the *General Form PDE* module. The coupled set of partial differential equations (PDEs) was solved using the *General Form PDE* module, while the mass and momentum conservation equations were solved using the *Laminar Flow* module. Initial and boundary conditions were applied in both the *General Form PDE* module and the *Laminar Flow* module. The time-dependent solver used was the Multifrontal Massive Parallel Sparse Direct Solver (MUMPS) with a memory allocation factor of 1.2 and a pivot threshold of 0.1. A segregated solution approach was employed, in which the dependent variables were solved sequentially rather than simultaneously. The first step was to solve the evolution equations for the components of the $\mathbf{Q}$ tensor, followed by the solution of the velocity and pressure fields using the NS equations. This method was chosen to reduce computational complexity and memory requirements.

The components of the $\mathbf{Q}$ tensor can have negative eigenvalues; therefore, for the schematic representation of the liquid crystal (LC) microstructure, a new tensor $\mathbf{M} = \mathbf{Q} + \frac{1}{3}\mathbf{I}$ is used. More details about this representation can be found in Ref. 56. The governing equations and simulation setup used in this paper have been validated in several studies conducted by our research group.[15,16,25,54,56]

The geometries used for the simulations are presented in Figs. 1(a) and 1(b), with $L = 1$ m and $H = 0.2$ m. Mesh convergence was assessed by comparing centerline velocity profiles across three mesh resolutions—normal (1104 elements), fine (1653 elements), and finer (2701 elements)—using a custom mesh in COMSOL. Quantitative error analysis indicated that velocity discrepancies between successive refinements were below 0.07%, confirming numerical convergence. Time-step convergence was verified using a second-order backward differentiation formula (BDF) with adaptive time stepping and a relaxation factor. Sensitivity analysis was conducted by varying the time step size, maintaining a relative error tolerance of $10^{-6}$. Figure 1(c) presents the centerline velocity profiles for the different mesh resolutions. The final mesh consisted of 1653 and 2255 elements, respectively, with an average mesh quality of 1.0.

No-slip boundary conditions were applied to the surrounding boundaries: regions I and II for the channel geometry and regions I through VI for the sudden contraction geometry. Periodic boundary conditions were enforced at both the inlet and outlet, as summarized in Tables I and II. A parabolic velocity profile was imposed at the inlet,

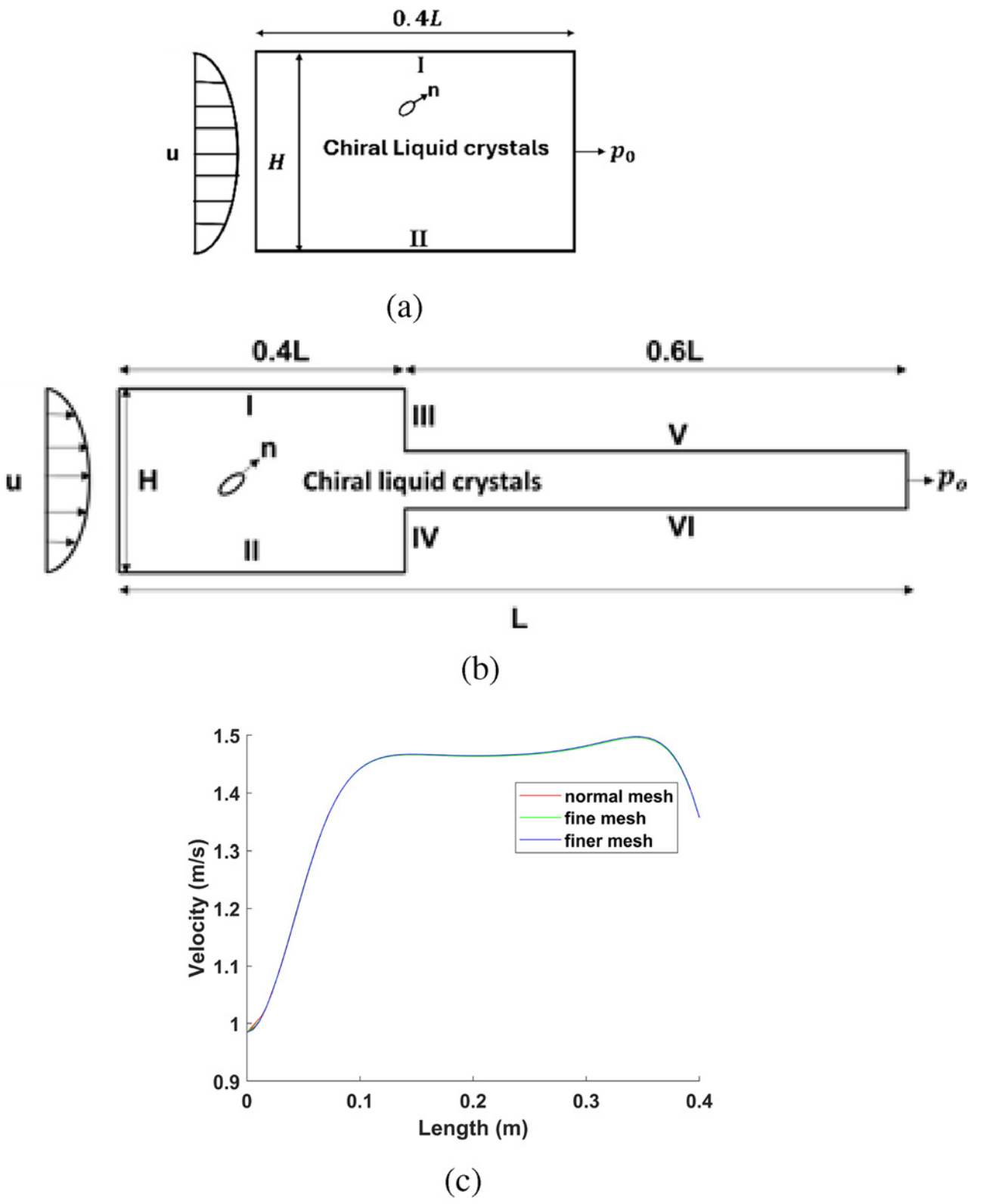

**FIG. 1.** Numerical setup and validation: (a) computational domain for the channel geometry, (b) computational domain for the sudden contraction geometry, and (c) centerline velocity profiles at different levels of mesh refinement.

**TABLE I.** Boundary conditions for velocity, $\boldsymbol{Q}$-tensor, and director field in the channel geometry.

| Location | $\boldsymbol{u}$ | $\boldsymbol{Q}^{\text{homeotropic}}$ | $\boldsymbol{n}^{\text{homeotropic}}$ |
|---|---|---|---|
| Inlet | Equation (26) | $\nabla \cdot \boldsymbol{Q} = 0$ | $\nabla \cdot \boldsymbol{n} = 0$ |
| Outlet | $\nabla \cdot \boldsymbol{u} = 0$ | $\nabla \cdot \boldsymbol{Q} = 0$ | $\nabla \cdot \boldsymbol{n} = 0$ |
| Boundary I and II | $\boldsymbol{u} = 0$ | $\left[-\frac{1}{3}, \frac{2}{3}, -\frac{1}{3}\right]$ | $(0, 1)$ |

as defined in Eq. (26), while zero pressure was specified at the outlet as follows:

$$u = u_0\left(1 - \left(\frac{y}{4H}\right)^2\right). \tag{26}$$

The maximum velocity, $u_0$, was set to 1 m/s. In the channel geometry, the director was oriented perpendicular to both boundaries. For the sudden contraction geometry, homeotropic boundary conditions (director perpendicular to the wall) were applied on boundaries I, II, V, and VI, while planar boundary conditions were used on boundaries III and IV. This configuration avoids discontinuities at the corners, which could otherwise lead to numerical artifacts.[25,32]

**TABLE II.** Boundary conditions for velocity, $\boldsymbol{Q}$ tensor, and director for the sudden contraction geometry.

| Location | $\boldsymbol{u}$ | $\boldsymbol{Q}$ | $\boldsymbol{n}$ |
|---|---|---|---|
| Inlet | Equation (26) | $\nabla \cdot \boldsymbol{Q} = 0$ | $\nabla \cdot \boldsymbol{n} = 0$ |
| Outlet | $\nabla \cdot \boldsymbol{u} = 0$ | $\nabla \cdot \boldsymbol{Q} = 0$ | $\nabla \cdot \boldsymbol{n} = 0$ |
| Boundary I–VI | $\boldsymbol{u} = 0$ | $\left[-\frac{1}{3}, \frac{2}{3}, -\frac{1}{3}\right]$ | $(0, 1)$ |

The initial conditions for the $\mathbf{Q}$ tensor were set to zero to introduce randomness, the initial flow velocity and pressure were also set to zero. The simulations were run for a total time of $t = 500$ s. An achiral case was used to verify the results against those reported by Fedorowicz and Prosser,[32] showing very good agreement. In all simulations, except for variations in $t$, Er, and $\Theta$, the following parameters were used: $\beta = 1$, $\varkappa = 4$, and $L^* = -0.3$. To ensure the predominant effect of viscous forces over inertia, the $\text{Re}_n$ was set to 1. However, since $\text{Re}_n$ was defined differently from its conventional form in this study, direct comparisons with standard $\text{Re}_n$ values are not appropriate. The energy ratio was selected as $R = 10^4$. The dimensionless viscosities were as follows: $\nu_1 = 0.5057$, $\nu_2 = 0.3138$, and $\nu_3 = -0.4383$.[16,25]

## IV. RESULTS AND DISCUSSION

Part 1 of this section presents the flow simulations of CLCs in a single channel, as this relatively simple geometry allows for better understanding of the effects of different parameters. Next, in part 2, simulation results of flows in channels with sudden contractions will be presented.

### A. Part 1: Flow of CLCs in a channel

This section presents the results on the flow of CLCs in a channel, focusing on how chirality, temporal variations, channel aspect ratio, and viscous effects influence the microstructure, as well as the coupling between the evolving microstructure and the flow.

#### *1. Chirality effect on the microstructure*

Figures 2(a)–2(f) present the molecular visualizations and $S$, of CLC flow in a channel under varying chiral strengths ($\Theta = 10, 20, 30, 40, 50,$ and $60$) for $\text{Er} = 10$. This value of Er was chosen to minimize viscous effects, ensuring that long-range elastic interactions dominate, thereby allowing a clearer observation of chiral effects and their role in structure formation. At low chiral strength ($\Theta = 10, 20$), the molecules exhibit uniform alignment, and as the chiral strength increases, two $\pi$-twist structures emerge [Fig. 2(c)]. They are characterized by defects at the upper, lower, and center regions of the structure.[59] These defects include $\tau^-$ and $\lambda^{(2+)}$ disclinations, with winding numbers of $s = -\frac{1}{2}$ and $s = +1$. Further increases in chiral strength induce bifurcation of the $\pi$-twist structures [Fig. 2(d)], driven by the coupling of flow and LC dynamics. This results in additional structures near the inlet region. At even higher chiral strengths [Figs. 2(e) and 2(f)], the defects organize into a hexagonal arrangement, marking the onset of polydomain structures. This arrangement is similar to the lattice-like configurations observed in BPs, particularly BP II, where double-twist cylinders are regularly arranged in a three-dimensional cubic lattice. Although classical BPs are typically observed in the bulk and under equilibrium conditions, the emergence of

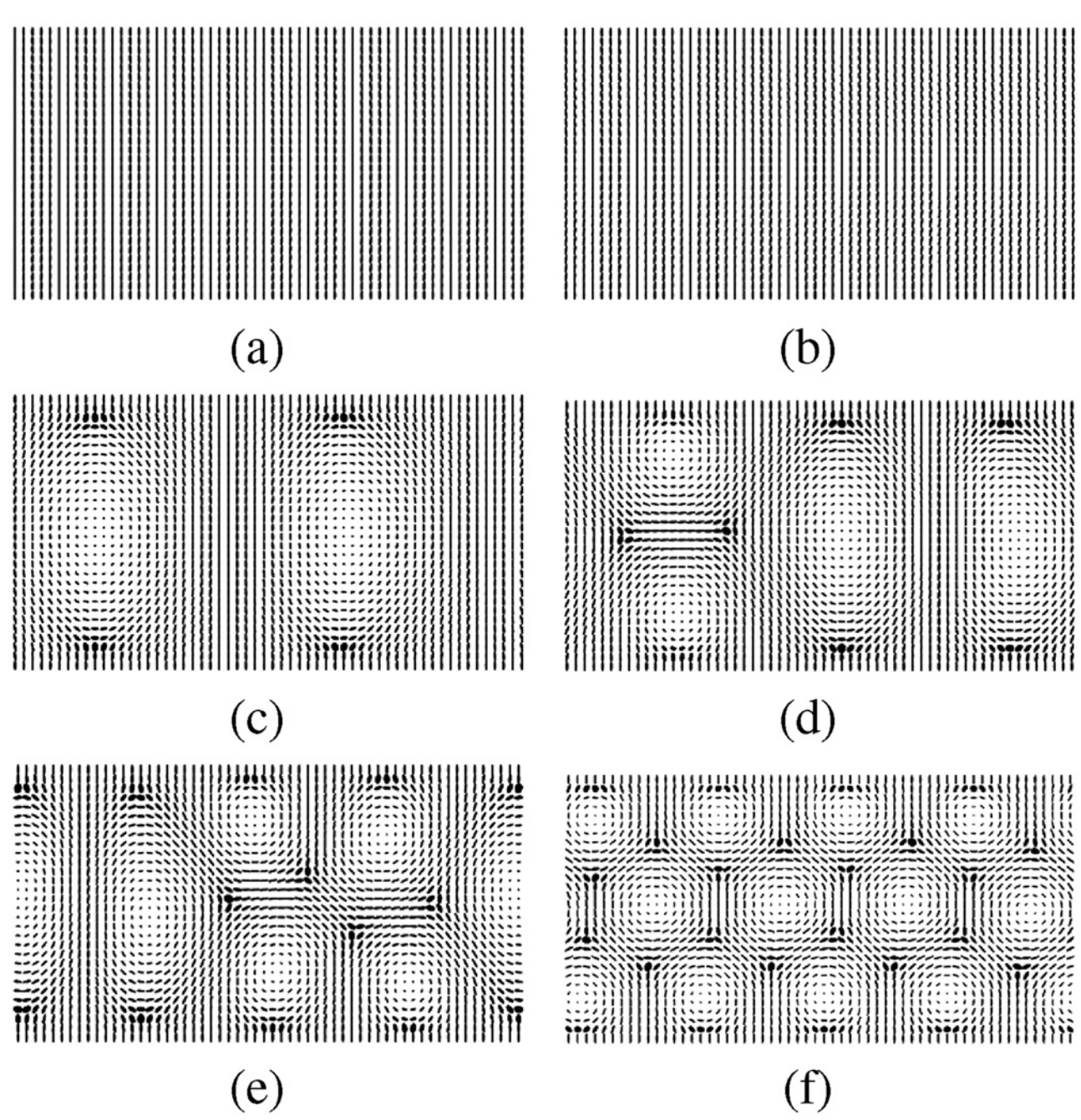

**FIG. 2.** Microstructure visualizations for Er = 10 and (a) $\Theta = 10$, (b) $\Theta = 20$, (c) $\Theta = 30$, (d) $\Theta = 40$, (e) $\Theta = 50$, and (f) $\Theta = 60$ at $t = 500$ s.

hexagonal patterns here under confinement and flow suggests a quasi-BPs-like regime adapted to the geometry.

Moreover, the tightly packed hexagonal defect arrangements at high $\Theta$ [Fig. 2(f)] consist of central $\lambda$-disclinations ($s = +1$) surrounded by six $\tau^-$-defects ($s = -\frac{1}{2}$), forming a net topological charge of zero. This configuration is consistent with previous reports of half-skyrmion structures in confined CLCs.[7,60] While we do not compute the skyrmion number explicitly, we refer to these textures as "skyrmion-like" based on their geometric similarity to these known topological configurations. Such skyrmion analogs have been reported in highly chiral CLCs, especially under confinement or external fields.[44,60]

This transition is further illustrated in Fig. 3, which captures the structural evolution from single-twist and double-twist configurations to the hexagonal pattern formed by three double twists. These observations are consistent with previous studies for other geometries.[15,25,61]

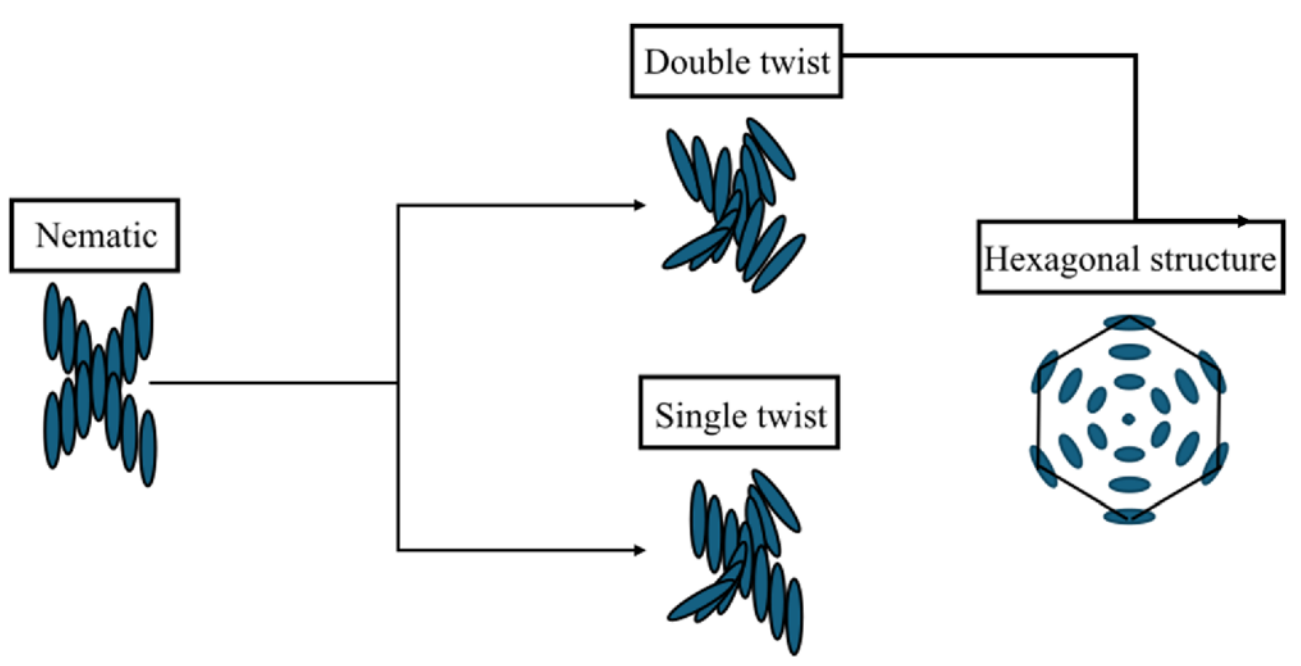

**FIG. 3.** Comparison of the nematic phase highlighting the structural differences between the single twist, double twist, and the hexagonal structure.

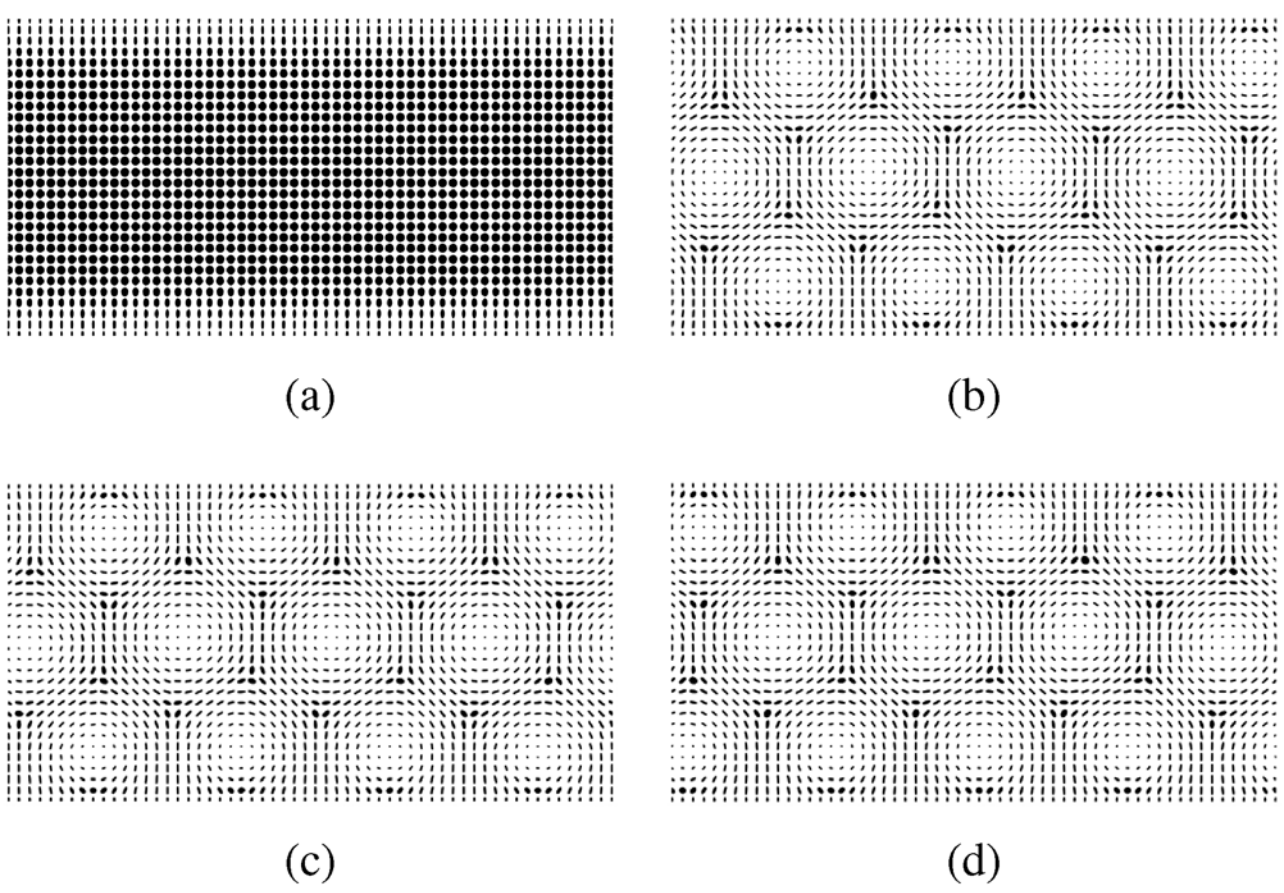

**FIG. 4.** Microstructure visualizations for Er = 10 and $\Theta = 60$ at (a) $t = 0$ s, (b) $t = 100$ s, (c) $t = 300$ s, and (d) $t = 500$ s.

### 2. Temporal effect on the microstructure

Figures 4(a)–4(d) show the evolution of the microstructure at high chiral strength ($\Theta = 60$) and low viscous effects over time. At $t = 0$ s, the system is isotropic as shown in Fig. 4(a), but as time progresses, a regular hexagonal defect lattice emerges [Figs. 4(b)–4(d)]. Each hexagonal unit cell is composed of a central $\lambda^{2+}$ disclination of strength $+1$, and six $\tau^-$ disclinations of strength $-\frac{1}{2}$ located at the vertices. Since each vertex is shared among three adjacent structures, only one-third of each vertex defect contributes to the charge of the cell. Thus, the net topological charge per hexagonal cell is zero. This is consistent with the theoretical condition as follows:

$$C_t = \sum \left( C_c + \frac{n_v}{3} C_v \right) = 0, \tag{27}$$

where $C_c$ represents the charge of the defect at the center, $n_v$ is the number of vertices, and $C_v$ denotes the charge of defects at the vertices of the unit cell, confirming that the structure forms a stable defect lattice.[61] Despite the emergence of long-range order, some polydomain regions retain non-uniform local charge distributions (at the walls), indicating a quasi-steady-state regime.[15,25,54,62] Notably, the hexagonal domains persist and move with the flow, suggesting the dynamic stability of defects.

### 3. Impact of the channel aspect ratio on the structure formation

Figures 5(a)–5(d) illustrate molecular visualizations and $S$ contours at different values of $\Theta$ (0, 20, 40, and 60, respectively) for different aspect ratios. At an aspect ratio $(H/L) = 0.25$, increasing the chiral strength causes a transition from achiral structures to $\pi$-twists and eventually to polydomain formations. These $\pi$-twists undergo further twisting as chiral effects dominate, leading to a reduction in pitch length. This behavior arises from the competition between the long-range elasticity of the nematic phase and the chiral torque, which progressively distorts the homeotropic molecular alignment.

A similar trend is observed with increasing aspect ratio $(H/L)$, as shown in Figs. 5(e)–5(h). Structural transitions occur earlier at $(H/L) = 0.375$ compared to 0.25, with the effect becoming even more pronounced at 0.5 [Figs. 5(i)–5(l)]. A higher aspect ratio provides lower effect of the long-range elasticity and this, in turn, enables the formation of more complex structures that are otherwise suppressed at lower values of $(H/L)$. Hence, the rise in $S$ with increasing aspect ratio and $\Theta$ reflects the system's capacity to form highly ordered, yet topologically rich structures. Nematic-like domains coexist with chiral-induced textures, such as $\pi$-twists and hexagonal arrangements, as shown in Fig. 5. These configurations are determined by the balance between confinement, elasticity, and chirality. Notably, the emergence of hexagonally arranged $\pi$-twists at higher aspect ratios and chiral strengths displays a strong resemblance to structural patterns observed in cholesteric BPs.

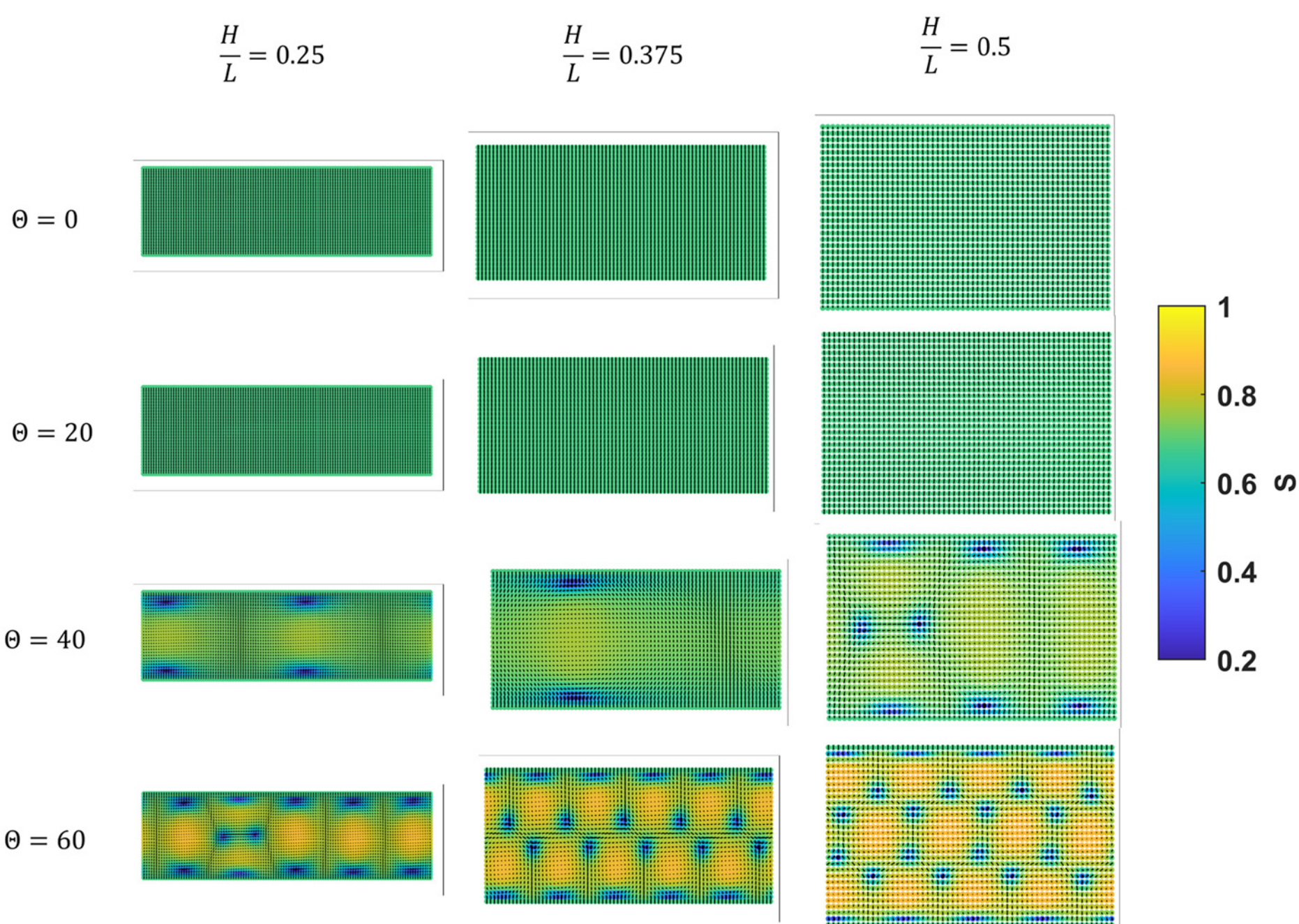

**FIG. 5.** Molecular visualizations and S contour plots for different chiral strengths ($\Theta$) and aspect ratios.

Figures 6 and 7 further illustrate this emergent complexity. Figure 6 shows the number of $\tau^-$ defects as a function of the chiral strength, $\Theta$, for different aspect ratios $(H/L)$. This highlights the role of geometric confinement in enhancing defect formation, as larger gaps reduce long-range elastic constraints and allow more defect formation. Figure 7 shows the increase in the number of defects for $(H/L = 0.5)$ at $\Theta = 60$. The progression from nematic order to $\pi$-twists, polydomain textures, and finally to hexagonal structures is presented. This result shows the connection between the aspect ratio, chiral effect, and long-range elastic effects.

#### 4. *Viscous effect on the microstructure*

In this section, the Er, was varied between 10 and 5000 for $\Theta = 60$, which is a high enough chiral strength to allow the formation of hexagonal structures. At low Er, distinct rows of chiral structures appear, as shown in Fig. 8(a). As Er increases, corresponding to higher viscous effects, these structures progressively transition from hexagonal to an irregular polydomain, followed by a region of low-$S$ line defects Figs. 8(b)–8(f). With increasing Er, polydomain structures start aligning with the flow direction [Fig. 8(b)]. At Er = 200 Fig. 8(c), hexagonal structures are partially broken. At higher Er, polydomain structures continue to diminish, eventually resulting in an achiral, flow-aligned monodomain Figs. 8(d)–8(f).

This evolution is reflected in the $S$: regions of high $S$ indicate uniform alignment, while low $S$ marks defect zones. These findings demonstrate that increasing viscous effects lead to defect annihilation and molecular alignment, driven by stronger viscous forces acting upon chirality and long-range elasticity. This structural evolution aligns with

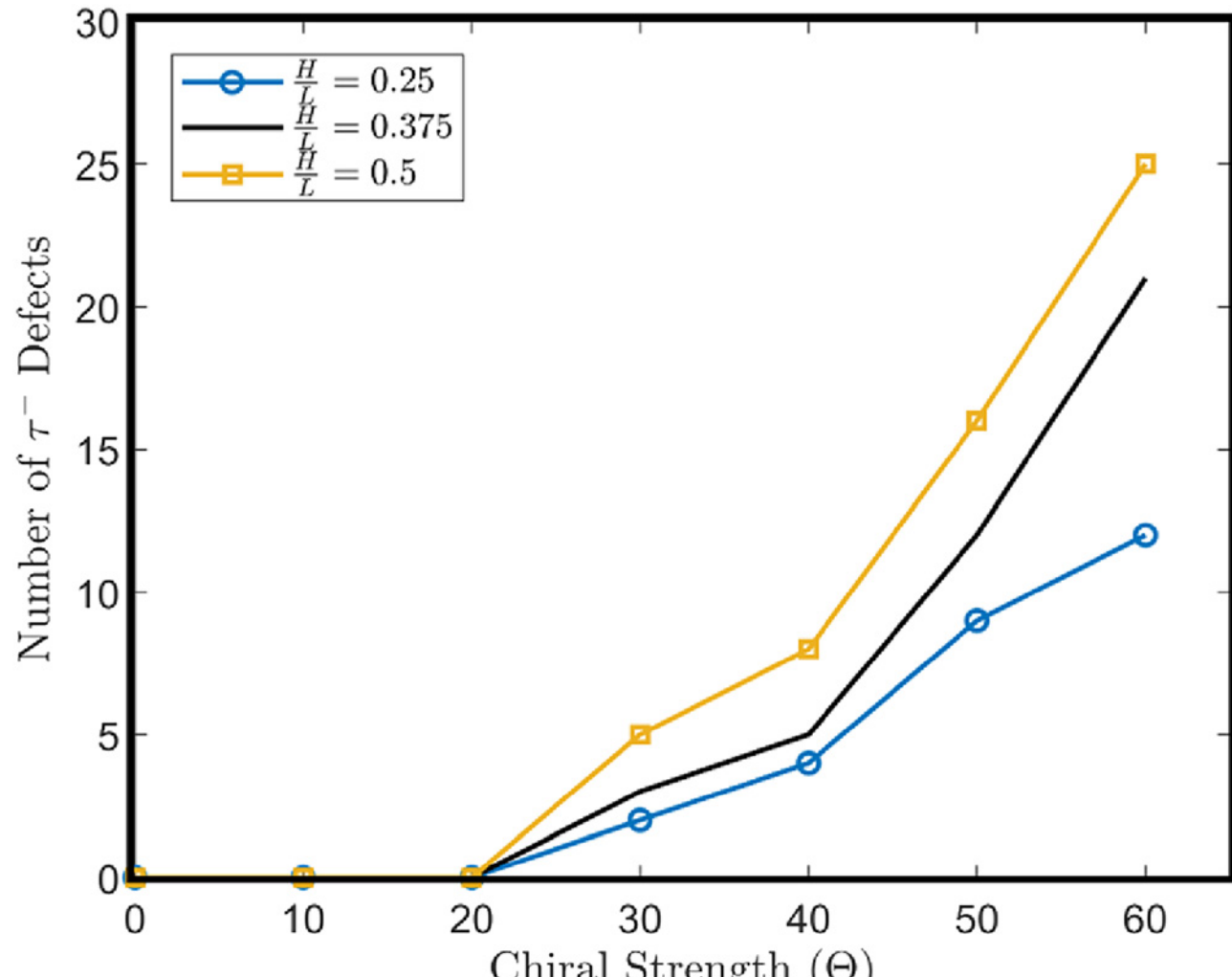

**FIG. 6.** Number of $\tau^-$ defects for different $(H/L)$ at $\Theta = 60$ with Er = 10 and $t = 500$ s.

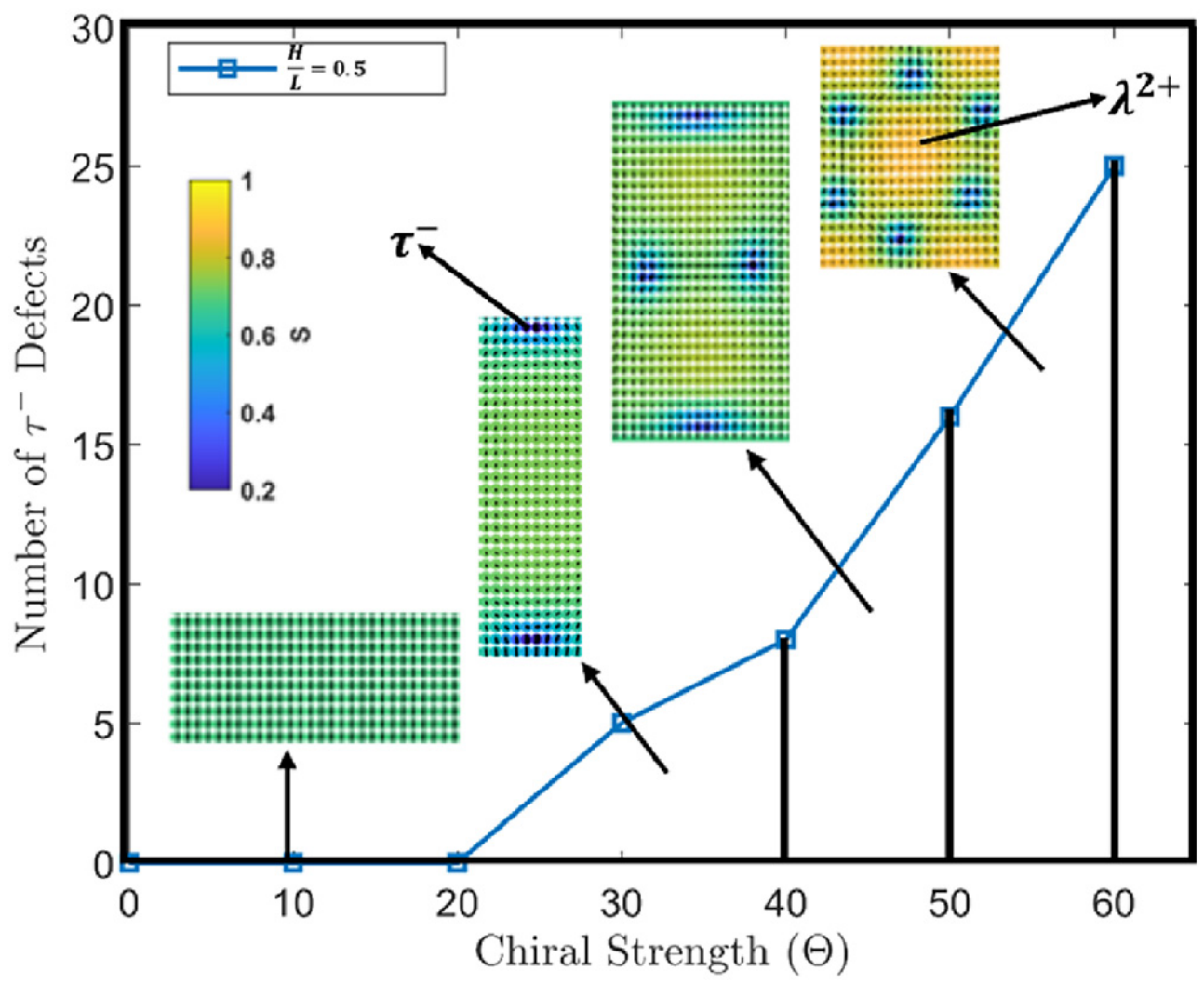

**FIG. 7.** Number of $\tau^-$ defects at $(H/L) = 0.5$ for $\Theta = 60$ at $Er = 10$ and $t = 500$ s. The zoomed-in section illustrates the corresponding structural transitions, progressing from nematic order to $\pi$-twists, then to polydomain configurations, and ultimately to hexagonal defect structures.

prior observations that viscous effects disrupt cholesteric twisting, leading to flow-induced unwinding.[63]

#### 5. Coupled effect of microstructure and the flow

Figures 9(a)–9(d) show the streamlines and velocity contours at different Er values ranging from 10 to 500. Long-range elastic effects become less pronounced at higher Er values as the impact of the boundary conditions is confined to a smaller spatial extent. At low Er, as shown in Fig. 9(a), the streamlines appear irregular due to the chiral structure formed, as shown in Fig. 8(a). Increasing the viscous effects to Er = 100, as shown in Fig. 9(b), reduces this non-uniformity, though some vortices remain at the walls and corners. At Er = 200, the number of small vortices formed at the wall decreases. At higher Er values (Er = 500), as shown in Fig. 9(d), a monodomain forms with no vortices present at the walls and corners. A more detailed analysis of the irregular flow regime observed at low Er will be presented in part 2 of this paper, focusing on a sudden contraction geometry, where such effects are further amplified.

### B. Part 2: Flow of CLCs in a sudden contraction geometry

This part presents the chiral and viscous effects on the microstructure, as well as how the microstructural changes affect the overall

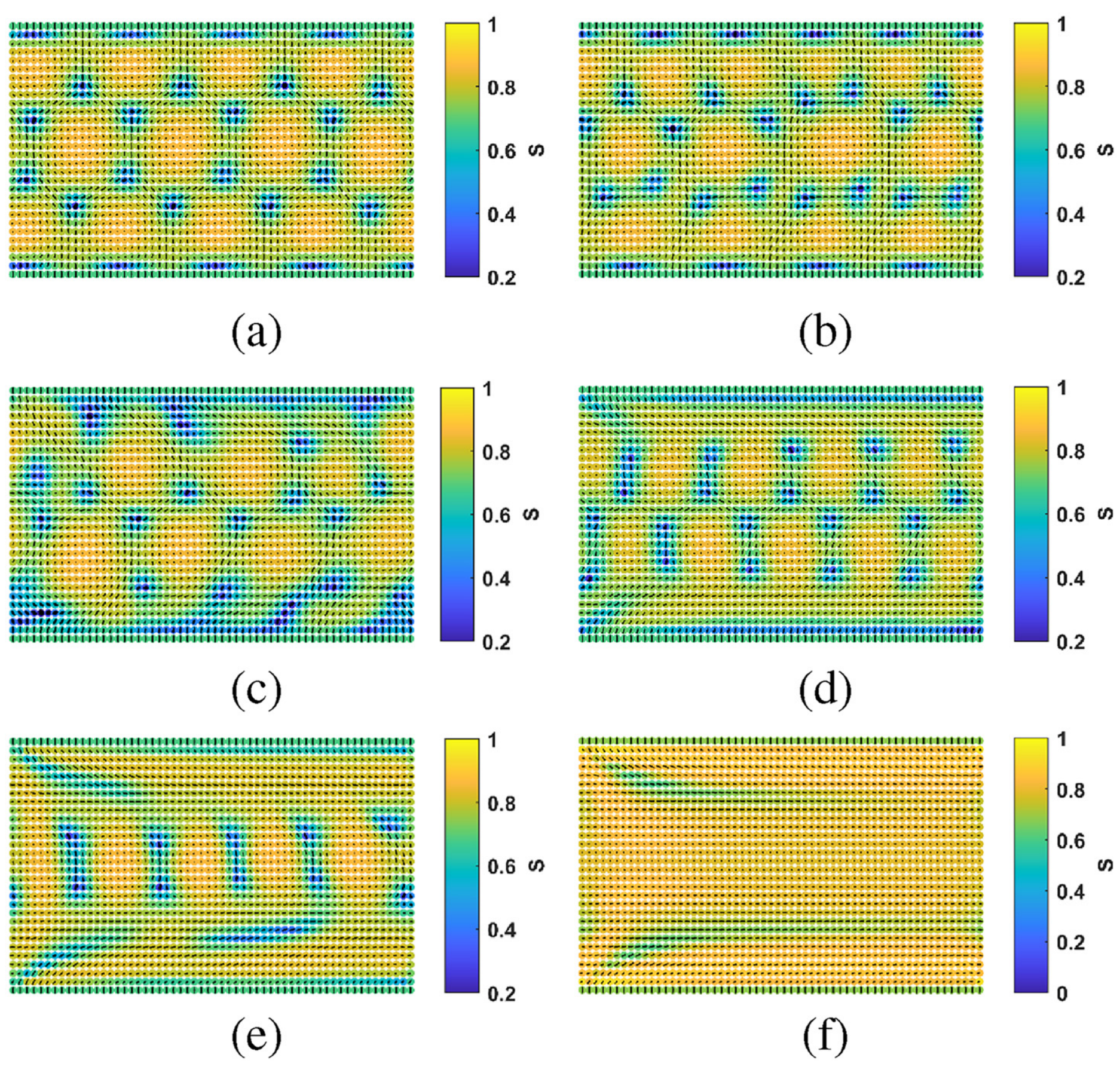

**FIG. 8.** Microstructure visualizations and S contour for $\Theta = 60$ and (a) Er = 10, (b) Er = 100, (c) Er = 200, (d) Er = 500, (e) Er = 1500, and (f) Er = 5000 at $t = 500$ s.

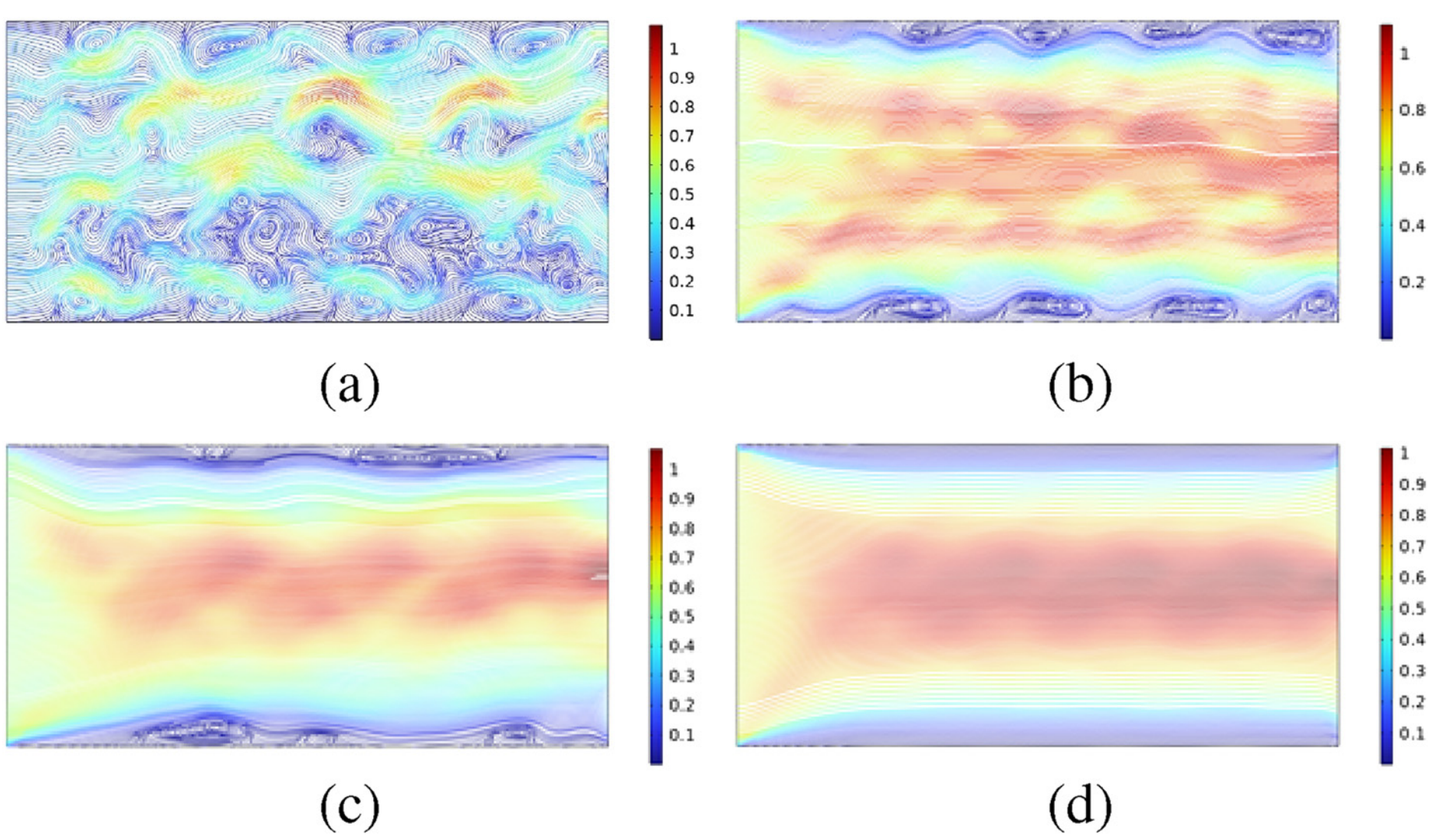

**FIG. 9.** Streamlines and velocity contours for $\Theta = 60$ and (a) Er $= 10$, (b) Er $= 100$, (c) Er $= 200$, and (d) Er $= 500$, at $t = 100$ s.

CLC flow behavior. This approach enables the investigation of the combined effects of geometry and CLC dynamics in a unified system. The aspect ratio $(H/L)$ is $\frac{1}{2}$ at the pre-contraction region and $\frac{1}{12}$ at the post-contraction region.

#### 1. Chiral effect on the microstructure

The formation of chiral structures in the sudden contraction geometry follows a similar progression as in the channel, but with additional effects induced by the contraction. Unlike the channel case, the localized velocity gradients in the pre-contraction region alter the molecular alignment and cause the emergence of structures. Figures 10(a)–10(d) illustrate the microstructure visualization and the S. The visualizations demonstrate the trend of increasing chiral strength $\Theta$ that leads to the emergence of multiple polydomain structures within the system. Specifically, at $\Theta = 20$ [Fig. 10(b)], a single lattice of $\pi$-twists appears at the top and bottom of the pre-contraction region, whereas in the channel, as shown in Fig. 8, these structures emerge at a higher chiral strength.

As $\Theta$ increases to 40 [Fig. 10(c)], additional rows of chiral structures are formed, which are separated by nematic regions and defects with strength $s = \frac{1}{2}$ at the corners of each structure. At $\Theta = 60$ [Fig. 10(d)], these structures are observed to self-organize into a hexagonal pattern, with nematic-like structures separating each lattice. Compared to the channel, where three double $\pi$-twists are the precursor of the formation of hexagonal domains, the contraction enhances disorder near the boundaries due to the boundary condition at the walls perpendicular to the channel walls at the contraction, which is in the direction of the flow. The resulting hexagonal arrays resemble BP-like lattices, where double-twist structures are stabilized under flow.

- Although the contraction geometry distorts the microstructure, it does not cause significant local disorder at $\Theta = 0$, but increasing chiral strength destabilizes the microstructure and leads to defect cores near the contraction edges, showing that chirality drives defect nucleation while geometry and anchoring influence their distribution.
- In the post-contraction region, chiral structures appear only at higher $\Theta$, accompanied by uniform polydomain regions along the flow path, while $\tau^{-}$ defects localize near the upper and lower boundaries of each structure. This observation aligns with prior research on confined geometries, where strong boundary effects induce localized defects and influence molecular alignment.[27,64,65] The increased defect density at higher $\Theta$ suggests a stronger interplay between chiral effects and flow-induced stresses in this region.

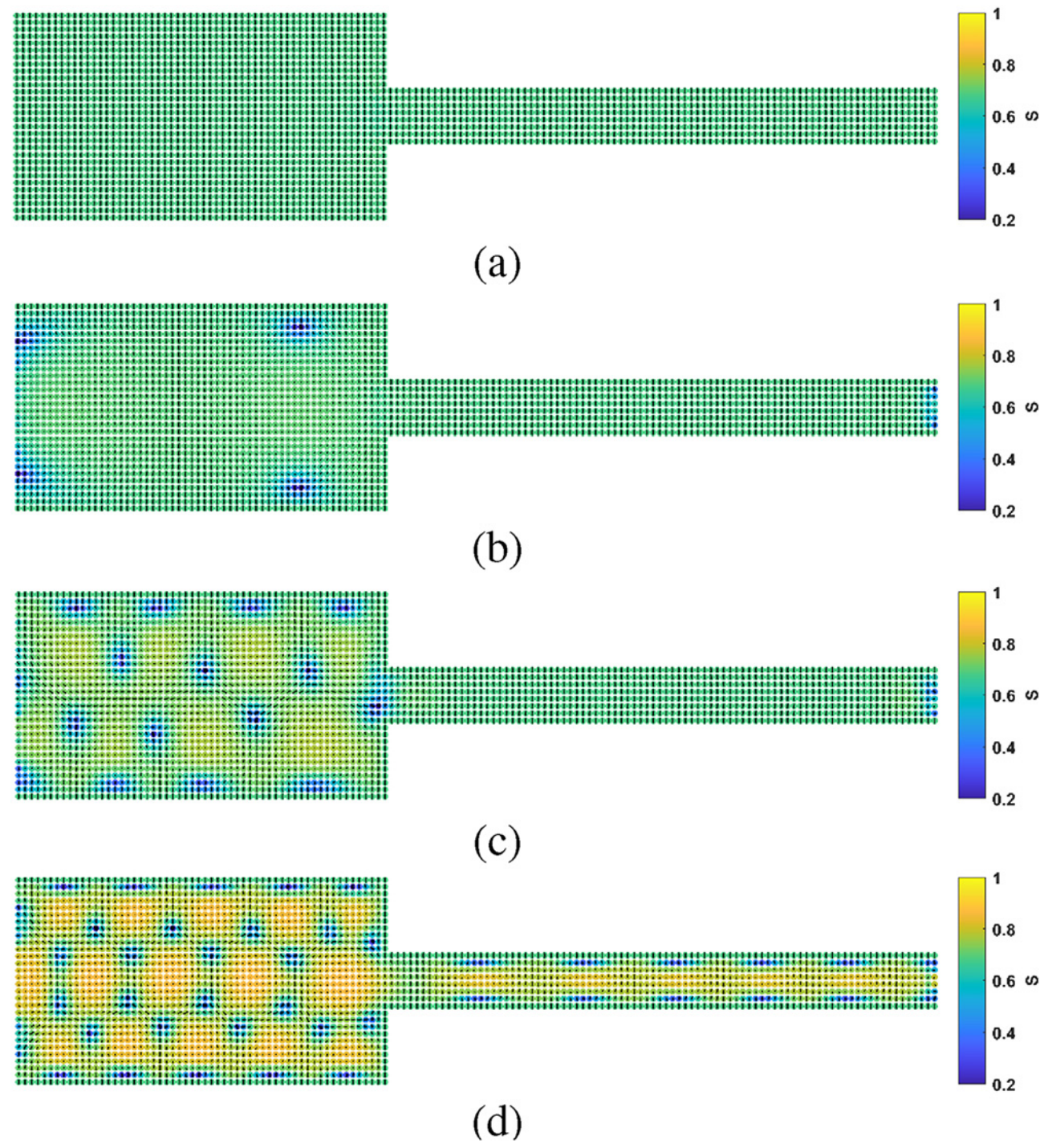

**FIG. 10.** Microstructure visualizations and $S$ contours for Er $= 10$ and (a) $\Theta = 0$, (b) $\Theta = 20$, (c) $\Theta = 40$, and (d) $\Theta = 60$ at $t = 500$ s.

### 2. Viscous effect on the microstructure

Figures 11(a)–11(e) present the molecular visualizations for Er ranging from 10 to 1000 at $\Theta = 60$. At lower Er values [Figs. 11(a)–11(c)], polydomain structures with defect zones persist in the pre-contraction region and gradually deform as Er increases. At $\mathrm{Er} \geq 500$ [Figs. 11(d) and 11(e)], these structures vanish, and flow-induced monodomains emerge. In the post-contraction region, low-$S$ regions initially align with $\tau^-$ disclinations but diminish with increasing Er, resulting in uniform molecular alignment. Due to variations in the channel aspect ratio, Er values are not directly comparable between the pre- and post-contraction regions. As Er increases, the structures gradually disappear, leading to higher-$S$ regions and enhanced flow alignment in the post-contraction channel. The contraction reduces the $H/L$, making the actual Er in the pre-contraction region higher than in the post-contraction region. This explains the greater number of chiral structure rows in the pre-contraction region. Beyond geometry, variations in velocity and shear rate further influence the viscous effects on either side of the contraction. Additionally, long-range elastic effects are more pronounced in the post-contraction region due to anchoring effects.

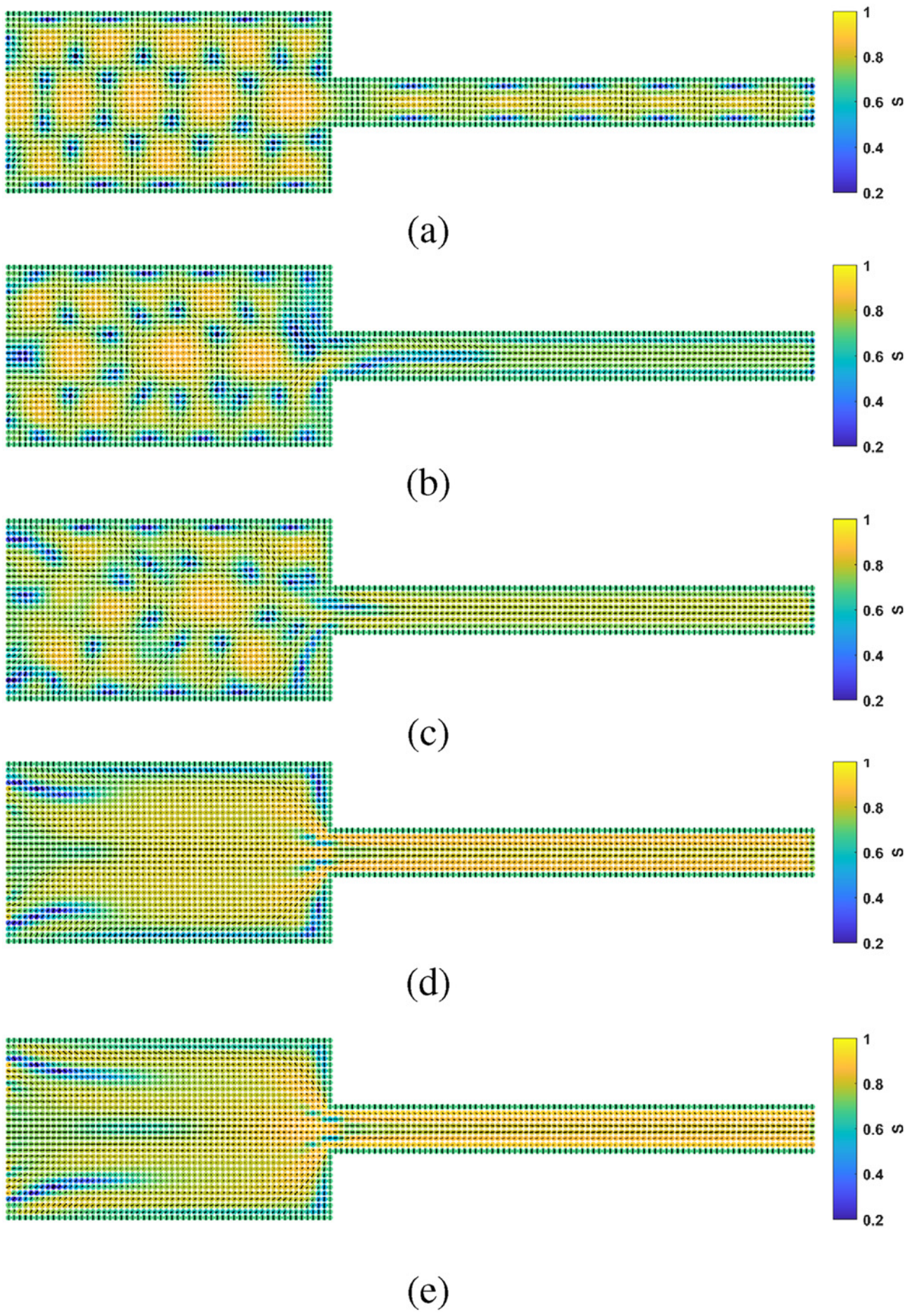

FIG. 11. Microstructure visualizations and $S$ contours for $\Theta = 60$ and (a) $\mathrm{Er} = 10$, (b) $\mathrm{Er} = 100$, (c) $\mathrm{Er} = 200$, (d) $\mathrm{Er} = 500$, and (e) $\mathrm{Er} = 1000$, at $t = 500\,\mathrm{s}$.

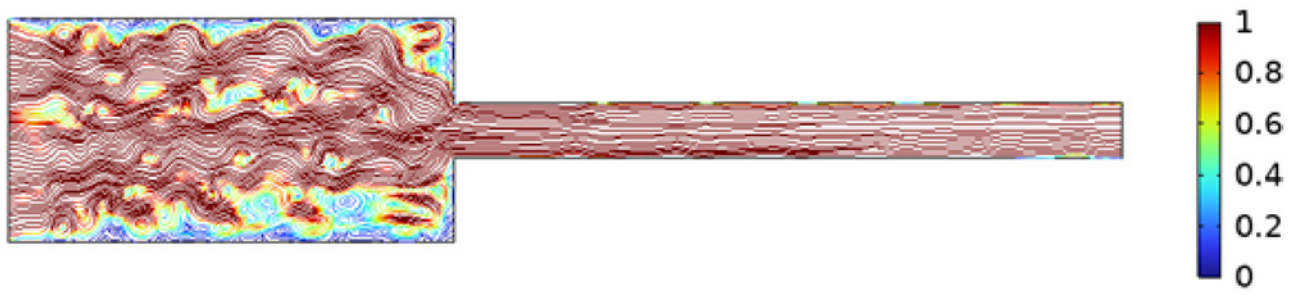

FIG. 12. Streamlines and velocity contours for $\Theta = 60$ and $\mathrm{Er} = 10$ at $t = 500\,\mathrm{s}$.

### 3. Coupled effect of microstructure and the flow

Figure 12 presents streamlines and velocity contours for $\Theta = 60$ and $Er = 10$ at $t = 100$ s. The flow field shows significant disturbance in the pre-contraction region compared to the post-contraction channel. This contrast is attributed to the difference in aspect ratio, where the wider geometry before the contraction allows for more complex flow patterns and instabilities, while in the narrower channel, the achiral microstructure stabilizes the flow.

At this low viscous effect, the influence of chirality is predominant, leading to the formation of a irregular flow in the pre-contraction region. Compared to the channel, increased disorder is observed due to the sudden contraction geometry, which enhances flow disturbances. Since the flow dynamics at higher $Er$ values follow the same trends as in the channel.

Vortex formation plays a crucial role in shaping the flow–microstructure coupling. The $\tau^-$ defects observed at the vortices of the hexagonal domains in Fig. 10(d) appear to arise from and reinforce local vortices, forming in regions of low velocity and low S, particularly at the pre-contraction where flow disturbances intensify. This suggests a coupled interaction, where the cholesteric microstructure induces vortices that, in turn, influence defect formation. These vortices arise due to the interplay between chiral and viscous effects, especially in regions subjected to abrupt reductions in cross-sectional area. The correlation between low-velocity regions, indicated by the streamlines in Fig. 12, and the presence of defects aligns with previous studies.[62] To quantify the localized disturbances within the flow, a cross-sectional plane intersecting the defect site—indicated by the black line in Fig. 13—was analyzed. Along this plane, the velocity magnitude and S were computed. The results, shown in Fig. 14, reveal a strong spatial correlation between regions of low velocity and areas of reduced S. This observation is consistent with the flow–structure coupling in CLCs, where topological defects disrupt molecular alignment, locally suppress momentum transfer, and lead to reduced flow coherence.

Further insight into the dynamics of the pre-contraction region is provided in Fig. 15(a), which shows the temporal evolution of velocity fluctuations at three representative locations along the flow domain: the inlet $(0, 0.1)$, the midpoint of the pre-contraction $(0.2, 0.1)$, and the contraction point $(0.4, 0.1)$. At the inlet, the velocity remains steady with minimal fluctuations, indicative of a stable flow regime and a well-aligned molecular field with negligible perturbations in the CLCs. In contrast, the midpoint region exhibits pronounced fluctuations, reflecting a disturbed state driven by the interplay between the flow and the CLCs microstructure. This disturbance stems from elastic frustration caused by the competing effects of intrinsic twist elasticity and flow-induced deformation as the CLCs approach the contraction. Within the contraction, the velocity increases sharply, but fluctuations are significantly reduced, suggesting that the strong extensional flow stabilizes the molecular configuration, suppressing distortions and restoring order. This behavior is further supported by the probability density functions (PDFs) shown in Fig. 15(b): the inlet exhibits the narrowest and most sharply peaked velocity fluctuation distribution, indicating highly stable flow. Although the contraction region is more geometrically confined, its distribution is broader than that of the inlet, suggesting slightly increased fluctuations. In contrast, the midpoint displays the broadest distribution, reflecting substantial velocity variability and microstructural rearrangement in this pre-contraction zone.[50]

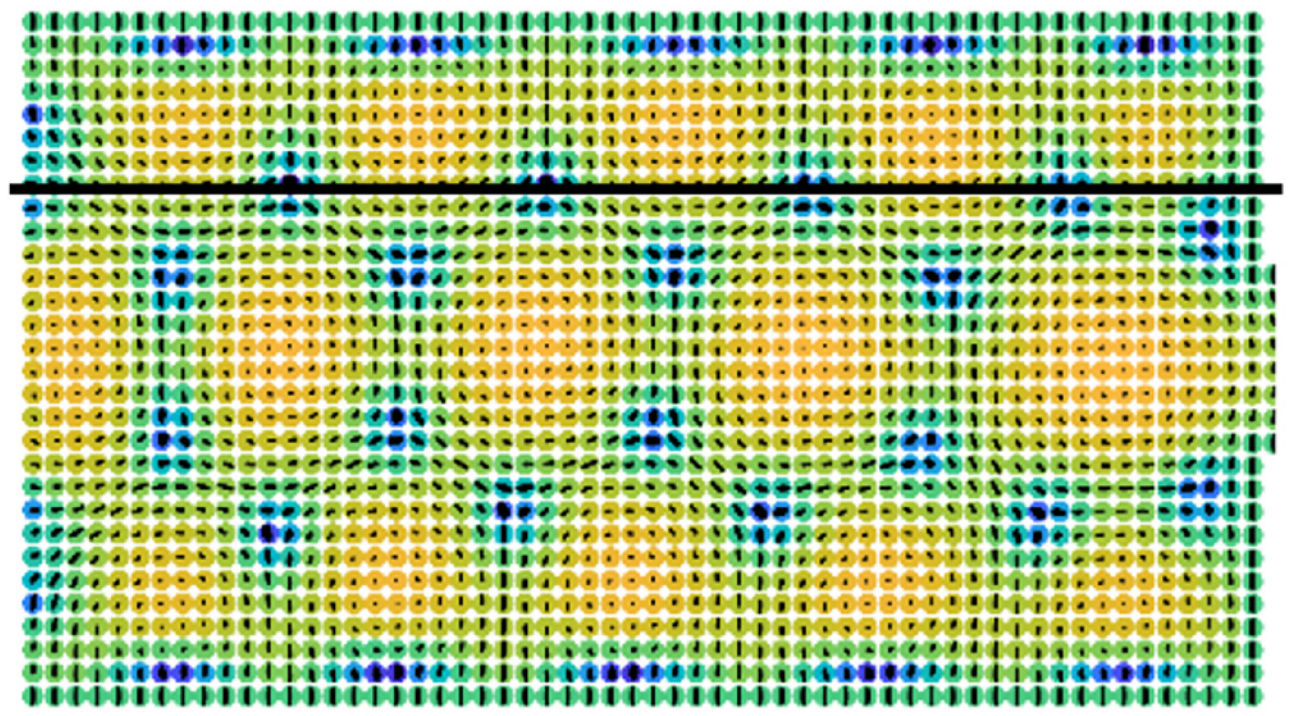

**FIG. 13.** Microstructure visualizations and S contours for $\Theta = 60$, $\mathrm{Er} = 10$, and $t = 500$ s in the precontraction region. The black line indicates the plane passing through the defect site.

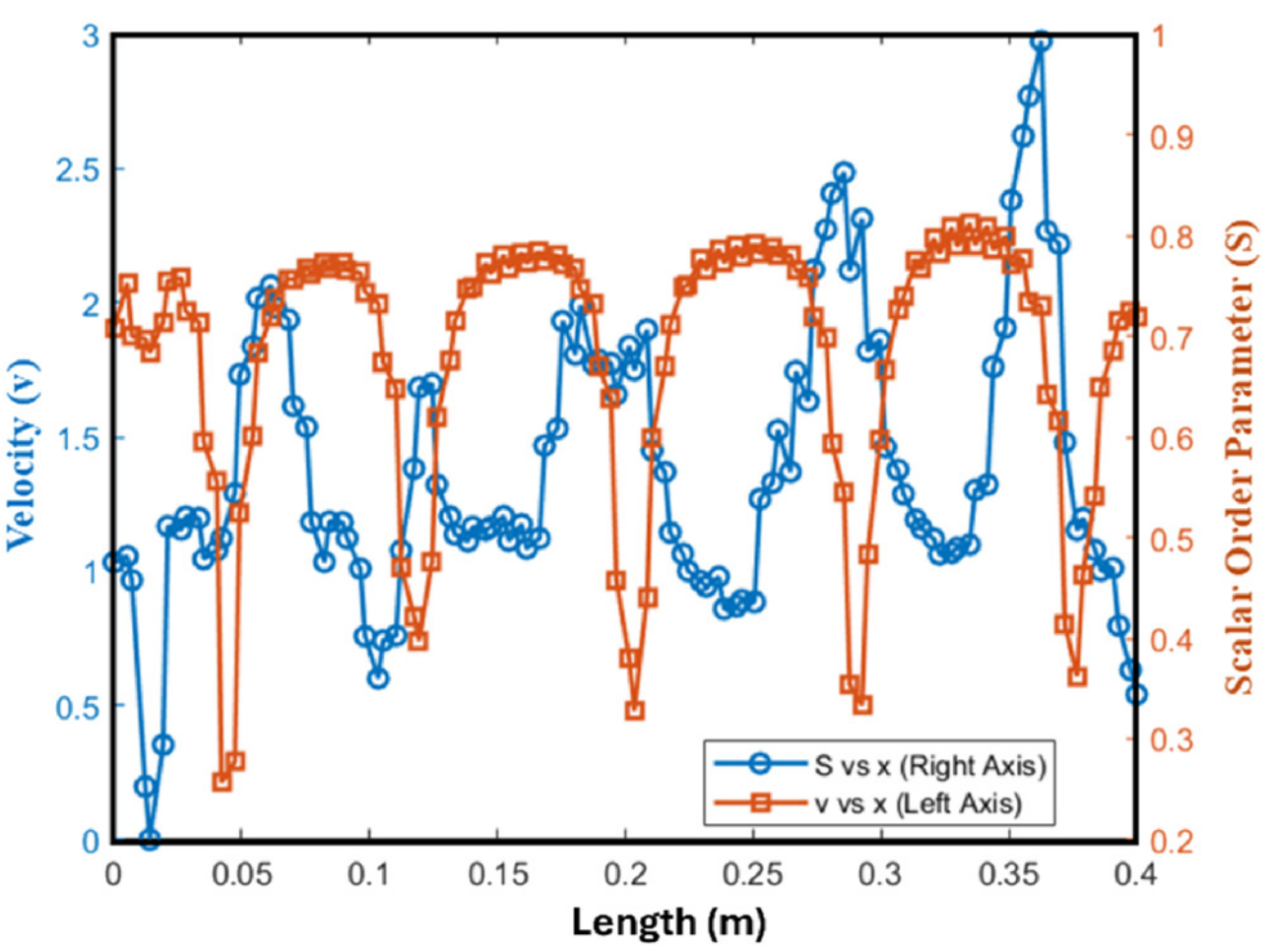

**FIG. 14.** Velocity magnitude and S plotted against the pre-contraction length for $\mathrm{Er} = 10$ and $\Theta = 60$ at $t = 500$ s.

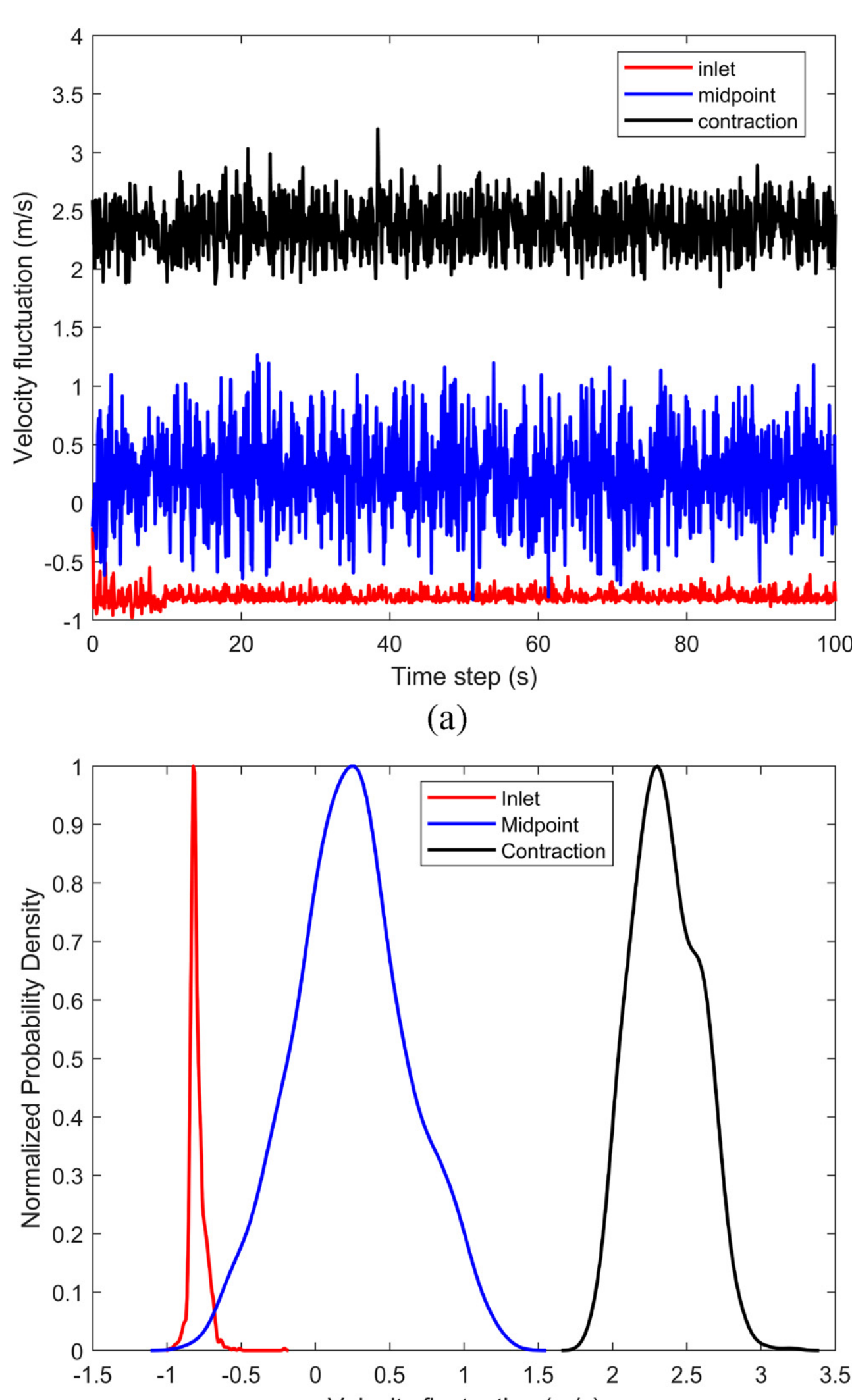

**FIG. 15.** (a) Temporal evolution of velocity fluctuations at three locations along the flow domain: the inlet point $(0, 0.1)$, the midpoint of the pre-contraction point $(0.2, 0.1)$, and the contraction point $(0.4, 0.1)$. Steady flow with minimal fluctuations is observed at the inlet and contraction zones, while the midpoint shows pronounced fluctuations due to strong flow–microstructure coupling. (b) Corresponding normalized PDFs of the velocity magnitudes at the three locations. Narrow, sharply peaked PDFs at the inlet reflect stable flow regimes, whereas the broader distribution at the midpoint and contraction indicates significant variability and molecular rearrangement.

## V. CONCLUSION

This study uses the LdG framework to provide a comprehensive insight into the flow behavior and structural dynamics of CLCs within channel and sudden contraction geometries. The results show how chirality, viscous effects, and long range elasticity drive the evolution of microstructures, including different defects, polydomain configurations, and hexagonal structures.

Key findings revealed that increasing chiral strength leads to developing $\pi$-twist defects at low chiral strength and eventually forming hexagonal structures at higher chiral strength, consistent with BPs-like ordering under confinement. At sufficiently high chirality and aspect ratios, localized defect arrangements emerge that are structurally analogous to half-skyrmion configurations observed in confined cholesteric systems. Although the skyrmion number is not explicitly calculated, the defect arrangement mirrors known equilibrium patterns. These results indicate dynamic analogs of equilibrium phases formed and stabilized by flow–structure interactions. These observations open possibilities for defect engineering, where precise control over chirality enables the design of targeted defect architectures with potential applications in photonic materials, soft robotics, and responsive surfaces. Additionally, increase in geometry aspect ratio facilitated the emergence of complex structures due to more spatial space, while viscous effects prompted the annihilation of chiral structures and finally formation of flow-aligned domains. In sudden contraction geometries, the connection between flow dynamics and structural changes emphasized the importance of long-range elasticity, velocity gradients, and chiral strength in defining the CLC behavior. These results advance our understanding of CLC flow dynamics in confined domains, offering practical applications for advanced manufacturing processes like microfluidics and 3D printing, where structural integrity and defect minimization are critical.

Furthermore, the findings show the potential of CLCs in drug delivery systems, where controlled orientation and microstructural changes could optimize therapeutic agent release. Although this study does not model drug transport or biocompatibility directly, the observed flow-controlled structural organization in CLCs may have implications for microfluidic drug delivery systems. Prior experimental work (e.g., Refs. 9 and 10) has demonstrated that cholesteric ordering enhances drug solubility and enables controlled release through structural anisotropy and pitch-tunable diffusion barriers. The ability to modulate defect density and alignment via flow and confinement may, therefore, inform the design of responsive delivery platforms. Another finding is the emergence of irregular flow regimes characterized by pronounced spatial and temporal fluctuations at low Er. The irregular regions, marked by fluctuating defect motion and irregular velocity distributions, highlight the nonlinear interactions between elastic stresses and flow dynamics. These fluctuations were quantified through velocity deviation profiles, where irregular regions exhibited sharp peak in the probability distributions.

While the BP-like arrangements observed in our simulations arise from cholesteric ordering imposed through a chiral term, we acknowledge that long-range twisted structures could, in principle, emerge from alternative molecular arrangements, including non-cholesteric systems. Moreover, although the present model assumes constant material parameters, CLC phases are known to be highly temperature-sensitive, with properties, such as elastic constants, viscosities, and pitch length, varying significantly with temperature. Investigating these alternative structural origins and incorporating temperature-dependent behavior into future works are both promising directions, particularly for capturing thermally induced transitions and dynamic instabilities. Future work will also focus on a deeper understanding of irregular structural dynamics, defect interactions, and the mechanisms driving flow instability.

## ACKNOWLEDGMENTS

D.G. acknowledges the financial support from the Natural Sciences and Engineering Research Council (NSERC) of Canada, Discovery Grant. I.M. acknowledges the financial support from the UBC Ambassador/4YF Scholarship. The authors gratefully acknowledge WestGrid and the Digital Research Alliance of Canada for providing computational resources for this research.

## AUTHOR DECLARATIONS

### Conflict of Interest

The authors have no conflicts to disclose.

### Author Contributions

**Isreal Morawo:** Conceptualization (equal); Data curation (equal); Formal analysis (equal); Investigation (equal); Methodology (equal); Resources (equal); Software (equal); Validation (equal); Visualization (equal); Writing – original draft (equal); Writing – review & editing (equal). **Dana Grecov:** Conceptualization (equal); Funding acquisition (equal); Methodology (equal); Project administration (equal); Resources (equal); Software (equal); Supervision (equal); Validation (equal); Writing – review & editing (equal).

## DATA AVAILABILITY

The data that support the findings of this study are available from the corresponding author upon reasonable request.